\documentclass[aip,jcp,preprint]{revtex4-1}
\usepackage{mhchem}
\usepackage{graphicx}
\usepackage{amsmath,amssymb}
 
\begin{document}
\title{Surface and Interfacial Tensions of Hofmeister Electrolytes}

\author{Alexandre P. dos Santos}
\email{alexandre.pereira@ufrgs.br}
\affiliation{Instituto de F\'isica, Universidade Federal do Rio Grande do Sul, Caixa Postal 15051, CEP 91501-970, Porto Alegre, RS, Brazil.}

\author{Yan Levin}
\email{levin@if.ufrgs.br}
\affiliation{Instituto de F\'isica, Universidade Federal do Rio Grande do Sul, Caixa Postal 15051, CEP 91501-970, Porto Alegre, RS, Brazil.}

\begin{abstract}
We present a theory that is able to quantitatively account for the surface and the interfacial tensions of different electrolyte solutions. It is found that near the interface ions can be separated into two classes: the kosmotropes and the chaotropes.  While the kosmotropes remain hydrated near the interface and are repelled  from it, the chaotropes loose their hydration sheath and become adsorbed
to the surface.  The anionic adsorption is strongly correlated with the Jones-Dole viscosity B-coefficient.  Both hydration and polarizability must be taken into account  to obtain a quantitative agreement with the experiments. To calculate the excess interfacial tension of the oil-electrolyte interface the dispersion interactions must be included.  The theory can also be used to calculate the surface and the interfacial tensions of acid solutions, showing a strong tendency for the surface adsorption of the hydronium ion.
\end{abstract}
 
\maketitle

\section{Introduction.}

Understanding behavior of ions at the air-water and  oil-water interfaces should 
help us to understand how these
ions interact with proteins and colloidal particles.  
Over a hundred years
ago Hofmeister organized various electrolytes according to their ability to 
salt-out protein solutions.  The sequences of anions and cations, 
which now bare his name, have been also observed in the 
fields of science as diverse as the
biophysics,
biochemistry, electro-chemistry and colloidal 
science~\cite{KoKo96,JiLi04,PeWe10,PeOr10,DoLe11}. 

Although the bulk thermodynamics of electrolyte solutions is fairly
well understood~\cite{DeHu23},
we still know little about ionic behavior when the 
translational symmetry is broken~\cite{La17,Wa24,OnSa34,LeFl01,BoWi01}. 
The surface and the interfacial
tensions provide us with an indirect indication of ionic distribution near
the interface.  A long time ago 
Heydweiller~\cite{He10} observed that 
addition of salt increases the 
surface tension of the air-water interface. Heydweiller also noted that the  
effect of different electrolytes followed the
sequence found by Hofmeister some years earlier.
The increase of the surface tension by electrolytes was soon attributed to the ionic
depletion from the interfacial region~\cite{La17}. Wagner~\cite{Wa24}
and Onsager and Samaras~\cite{OnSa34} suggested that this was a consequence of the
charge induced on the dielectric interface separating water from air. 
The theory was able to quantitatively account for  the surface tension of sodium
chloride solution at very large dilutions, but failed for larger concentrations of 
electrolyte. Inclusion of ionic hydration  into the theory 
by Levin and Flores-Mena~\cite{LeFl01} extended its validity up to 
$1$M concentration.  However when the same theory 
was applied to study the surface tension of sodium iodide solutions it was
found that it predicts a qualitatively incorrect behavior  --- the surface
tension of \ce{NaI} was found to be larger than of \ce{NaCl}, contrary to experiment. 
The origin of this discrepancy was not clear.  
The fundamental insights, however, appeared soon after in the form of
polarizable force fields simulations~\cite{PeBe91,DaSm93,StBr99,JuTo02,JuTo06,Br08,HoHe09} and
experiments~\cite{MaPo91,Ga04,Gh05}.  The new simulations and experiments
demonstrated that it was possible for large halogen anions to become adsorbed
to the interface.  The physical mechanism of this adsorption, however, remained
unclear.  Bostr\"om et al.~\cite{BoWi01} suggested that the dispersion (van der Waals) interactions,
neglected within the Wagner-Onsager-Samaras (WOS) theory, were 
responsible for the ionic specificity.  This interesting suggestion, however, 
contradicts both experiments and simulations.  Dispersion forces are proportional to
the ionic polarizability.  Therefore, strong dispersion interactions between
ions and water should favor bulk solvation.  Since anions are much more
polarizable than cations, the dispersion interactions 
should keep these ions in the bulk, away
from the interface.  This means that a theory based on dispersion interactions 
will predict that
weakly polarizable cation should be adsorbed at the interface, 
contrary to what was found in experiments and simulations.    

A different theory was recently proposed by Levin~et~al.~\cite{Le09,LeDo09}.
These authors argued that the driving force behind the
adsorption of highly polarizable anions was due to the hydrophobic effect.  To solvate
an ion, a cavity must be created.  The cavity perturbs the hydrogen
bond network of water molecules, resulting in a free energy cost.  Clearly
if the ion moves towards the interface, the cavitational energy will diminish.
There is, however, an electrostatic self-energy penalty of exposing the ionic charge
to the low dielectric air environment.  For hard, non-polarizable, ions of WOS theory
the self-energy penalty completely overwhelms the gain in the hydrophobic free
energy, forcing these ions to remain in the bulk.  The situation is very different
for large polarizable anions.  When such ions move towards the interface,
their electronic charge distribution shifts so that it remains mostly hydrated.
This drastically diminishes the electrostatic self-energy penalty of having
such ions located at the interface.  A careful calculation shows that
for polarizable ions the electrostatic self-energy penalty becomes comparable to
the gain in the hydrophobic free energy resulting from moving an ion from the 
bulk to the surface~\cite{Le09}.

In this paper we will show how the ideas presented above can be used to calculate the
surface and the interfacial tensions of electrolytes and acid solutions, as well as their
electrostatic potential difference across the dielectric interface.

\section{The drop model.}

To perform the electrostatic calculations it is convenient to consider an electrolyte
solution inside a spherical water drop of radius $R$~\cite{HoTs03,LeDo09}.  
Here we will choose $R=300$~\AA\, which is sufficient large to avoid all
finite size effects, so that the excess surface tension calculated inside
a drop will be the same as the surface tension of an extended thermodynamic interface.
Outside the water drop is the low dielectric medium (air or oil) and the
interface at $r=R$ corresponds to the  
Gibbs dividing surface~(GDS). $N$ salt or acid ``molecules" are dissociated inside 
the drop, resulting in $N$ cations and $N$ anions of charge $+q$ and $-q$, respectively.  In the case of divalent
anions, for each anion of charge $-2q$ there will be $2$ cations of charge $+q$. 
The water and the external medium will be treated as dielectrics of permittivities $\epsilon_w$ and $\epsilon_o$, respectively. The 
Bjerrum length is defined as $\lambda_B=\beta
q^2/\epsilon_w$. 

The interfacial tensions are calculated by integrating the Gibbs adsorption isotherm
equation,
%%%%%%%%%%%%%%%%%%%%%%%%%%%%%%%%%%%%%%%%%%%
\begin{equation}\label{gaie}
{\rm d} \gamma=-\Gamma_+ {\rm d} \mu_+ - \Gamma_- {\rm d} \mu_- \ ,
\end{equation}
%%%%%%%%%%%%%%%%%%%%%%%%%%%%%%%%%%%%%%%%%%
where $\Gamma_{\pm}=\left[N - V \rho_{\pm}(0) \right]/S$ are the ionic excess
concentrations, $\mu_\pm$ are the chemical potentials, $\rho_{\pm}(0)$ are the
bulk concentrations, $S$ and $V$ are the surface and volume of the drop,
respectively. The bulk concentrations are
obtained from the numerical solution of the modified Poisson-Boltzmann (PB) equation:
%%%%%%%%%%%%%%%%%%%%%%%%%%%%%%%%%%%%%%%%%%%
\begin{eqnarray}\label{pb}
\nonumber
\nabla^2 \phi(r)&=&-\frac{4\pi q }{\epsilon_w} \left[\rho_+(r)-\rho_-(r)\right]  \ ,  \\
\rho_{\pm}(r)&=& A_{\pm} e^{\left[\mp \beta q \phi(r)-\beta U_{\pm}(r)\right]} \ ,  \\
\nonumber
A_{\pm}&=&N/\left[4\pi \int_0^{r_{max}} dr\ r^2 e^{\left[\mp \beta q \phi(r)-\beta U_{\pm}(r)\right]}\right] \ ,
\end{eqnarray}
%%%%%%%%%%%%%%%%%%%%%%%%%%%%%%%%%%%%%%%%%%%%
where $\phi(r)$ is the electrostatic potential, $\rho_\pm(r)$ are the ionic
concentrations, and $r_{max}$ is the maximum ionic distance from the center of the drop. For chaotropes 
$r_{max}=R+a$ and for kosmotropes $r_{max}=R-a$. The ion-interface interaction
potentials, $U_\pm(r)$, will be discussed in the following sections.  The
chemical potentials inside the drop are constant and within the   
PB approximation are given by $\beta \mu_\pm= \log{(\Lambda_\pm^3 \rho_{\pm}(0))}$, where
$\Lambda\pm$ are the thermal de Broglie wavelengths. 

\section{Air-water interface.}

When an ion moves close to the dielectric interface, there are two effects:
(1) the interface becomes polarized; and (2) there is a loss of 
solvation free energy arising from the imperfect screening of the ionic electric field
by the rest of electrolyte.  Both of these effects lead to a repulsive force from the interface.
The work necessary to bring an ion from  the bulk to a distance $z$ from the GDS is found to be~\cite{LeFl01,DoDi10} 
%%%%%%%%%%%%%%%%%%%%%%%%%%%%%%%%%%%%%%%%%%%%
\begin{equation}\label{Uim}
\beta U_{i}(z)=\beta W\frac{a}{z} \ e^{-2 \kappa (z-a)} \ ,
\end{equation}
%%%%%%%%%%%%%%%%%%%%%%%%%%%%%%%%%%%%%%%%%%%%
where $\kappa=\sqrt{8 \pi \lambda_B c_\pm}$ is the inverse Debye length and $a$
is the ionic radius. The contact value, $W$, is calculated by solving the Poisson equation
with the appropriate boundary conditions~\cite{LeFl01},
%%%%%%%%%%%%%%%%%%%%%%%%%%%%%%%%%%%%%%%%%%%%
\begin{equation}
\label{e2}
\beta W= \frac{\lambda_B}{2} \int_0^\infty dp \frac{p[ s \cosh(p a)-p \sinh(p
a)]}
{s[ s \cosh(p a)+ p \sinh(p a)]} \ ,
\end{equation}
%%%%%%%%%%%%%%%%%%%%%%%%%%%%%%%%%%%%%%%%%%%%
where $s=\sqrt{\kappa^2 + p^2}$.

For polarizable ions, Levin~\cite{Le09} calculated the variation in the
electrostatic self-energy
as an ion crosses the dielectric interface. The ion
was modeled as an imperfect conducting sphere of relative polarizability
$\alpha=\gamma/a^3$, where $\gamma$ is the absolute ionic polarizability
measured in \AA$^3$. The electrostatic self-energy of an ion whose center is at distance
$-a<z<a$ from the GDS is found to be
%%%%%%%%%%%%%%%%%%%%%%%%%%%%%%%%%%%%%%%%%%%%
\begin{equation}\label{Upol}
\beta U_{p}(z)= \frac{\lambda_B}{2 a}\left[\frac{\pi x^2}{\theta(z)}+\frac{\pi
[1-x]^2 \epsilon_w}{[\pi-\theta(z)]\epsilon_o}\right] + g \left[
x-\frac{1-cos[\theta(z)]}{2} \right]^2 \ ,
\end{equation}
%%%%%%%%%%%%%%%%%%%%%%%%%%%%%%%%%%%%%%%%%%%%
where $\theta(z)=\arccos[-z/a]$ and $g=(1-\alpha)/\alpha$. The fraction of
charge that remains hydrated $x$, is obtained by minimizing eqn~(\ref{Upol}),
%%%%%%%%%%%%%%%%%%%%%%%%%%%%%%%%%%%%%%%%%%%%
\begin{equation}
x(z)=\left[ \frac{\lambda_B \pi \epsilon_w}{a \epsilon_o
\left[\pi-\theta(z)\right]}+g [1-cos[\theta(z)]] \right] / \left[\frac{\lambda_B
\pi}{a \theta(z)} + \frac{\lambda_B \pi \epsilon_w}{a \epsilon_o
[\pi-\theta(z)]} +2 g \right] \ .
\end{equation}
%%%%%%%%%%%%%%%%%%%%%%%%%%%%%%%%%%%%%%%%%%%%

To solvate an ion in water requires creation of a cavity.  For small cavities
this hydrophobic free energy scales with the volume of the void~\cite{RaTr05}. 
When the ion crosses the GDS, the perturbation to the hydrogen
bond network diminishes, resulting in a thermodynamic force that drives the ion
towards the air-water interface. The cavitational potential energy is found to be
%%%%%%%%%%%%%%%%%%%%%%%%%%%%%%%%%%%%%%%%%%%
\begin{eqnarray}\label{cavpot}
\beta U_c(z)=\left\{
\begin{array}{l}
 \nu a^3 \text{ for } z \ge  a  \ , \\
 \frac{1}{4} \nu a^3  \left(\frac{z}{a}+1\right)^2 \left(2-\frac{z}{a}\right)
\text{ for } -a<z<a \ ,
\end{array}
\right.
\end{eqnarray}
%%%%%%%%%%%%%%%%%%%%%%%%%%%%%%%%%%%%%%%%%%%%
where $\nu \approx 0.3/$\AA$^3$, is obtained from the bulk simulations~\cite{RaTr05}.

The physical chemists have known for a long time that ions come in two 
categories:  structure-makers (kosmotropes) and structure-breakers (chaotropes).
The separation into these two classes is often based on the Jones-Dole~(JD)
viscosity B-coefficient~\cite{Ma09} which also correlates well with the ionic
enthalpy of hydration~\cite{Co97}.  In 1929 Jones and Dole observed that the relative
viscosity $\eta_r$ of an electrolyte solution is very well 
fit by a simple formula $\eta_r=1+A \sqrt{c}+B c$, where $c$ is the bulk concentration
of electrolyte.  The square root term is universal
and can be calculated using the Debye-H\"uckel-Onsager theory.  
On the other hand, the linear term
in $c$ is electrolyte specific. Ions with a positive
B coefficient~(kosmotropes) are supposed to organize water making
it ``more" viscous, while the ions with negative B are supposed to make 
water more disordered and ``less" viscous.  To what extent this physical picture is
realistic is not clear, and recent experiments suggest that the action of
ions on water molecules is much more local, not extending much beyond the first hydration
shell~\cite{OmKr03}. This view also fits well with the theory of surface tensions of electrolyte solutions that will be presented in this paper.
We find that kosmotropic ions remain strongly hydrated near
the interface and are repelled from it, while the chaotropic ions loose their
hydration sheath and as the result of their large polarizability 
become adsorbed to the interface.  

In the case of halides, the separations into kosmotropes and the chaotropes
correlates well with the ionic size.  Heavy halogen anions produce 
weak electric field that is not 
sufficient to bind the adjacent water molecules which dissociate
from the anion when it moves towards the  interface.  
This is the case for \ce{I-} and \ce{Br-}, whose 
JD viscosity B-coefficients are $-0.073$ and $-0.033$, respectively.  
On the other hand, the small 
fluoride anion has
large positive B-coefficient, $0.107$, signifying a strong interaction 
with the surrounding
water molecules. This ion should remain strongly hydrated near the interface. 
The chloride ion, with B-coefficient close to zero, $-0.005$, is on the
borderline between the two classes.
%In principle, all the discussed interactions potentials can be applied for all
%ions, but the situation is not so simple.
%Simulations\cite{JuTo02,JuTo06,HoHe09,Br08} and
%experiments~\cite{MaPo91,Gh05,Ga04} show that ionic sodium and fluoride are
%strongly repelled from the interface. We can argue that these ions have small
%bare ionic radius (high charge density), for instance their interaction with
%water molecules is very strong in a way that they never lose hydration, in other
%words, they never cross the GDS~\cite{LeDo09}. 

The total interaction potential
for strongly hydrated kosmotropes is dominated by the charge-image interaction,
$U_\pm(z)=U_i(z)$ and the hard core repulsion (at one hydrated radius) from the GDS. 
On the other hand, large chaotropic anions, such as 
iodide and bromide, are able to cross the GDS with relatively
small  electrostatic self-energy penalty, gaining the hydrophobic cavitational
free energy. For such chaotropic anions the total interaction potential is 
$U_-(z)=U_i(z)+U_p(z)+U_c(z)$. The radius~\cite{LaPi39} of 
\ce{I-} is $a=2.26$~\AA\ and its relative polarizability~\cite{PyPi92} is 
$\alpha=0.64$; for \ce{Br-},  $a=2.05$~\AA\ and
$\alpha=0.59$.

\begin{figure}[h]
\centering
  \includegraphics[height=8cm]{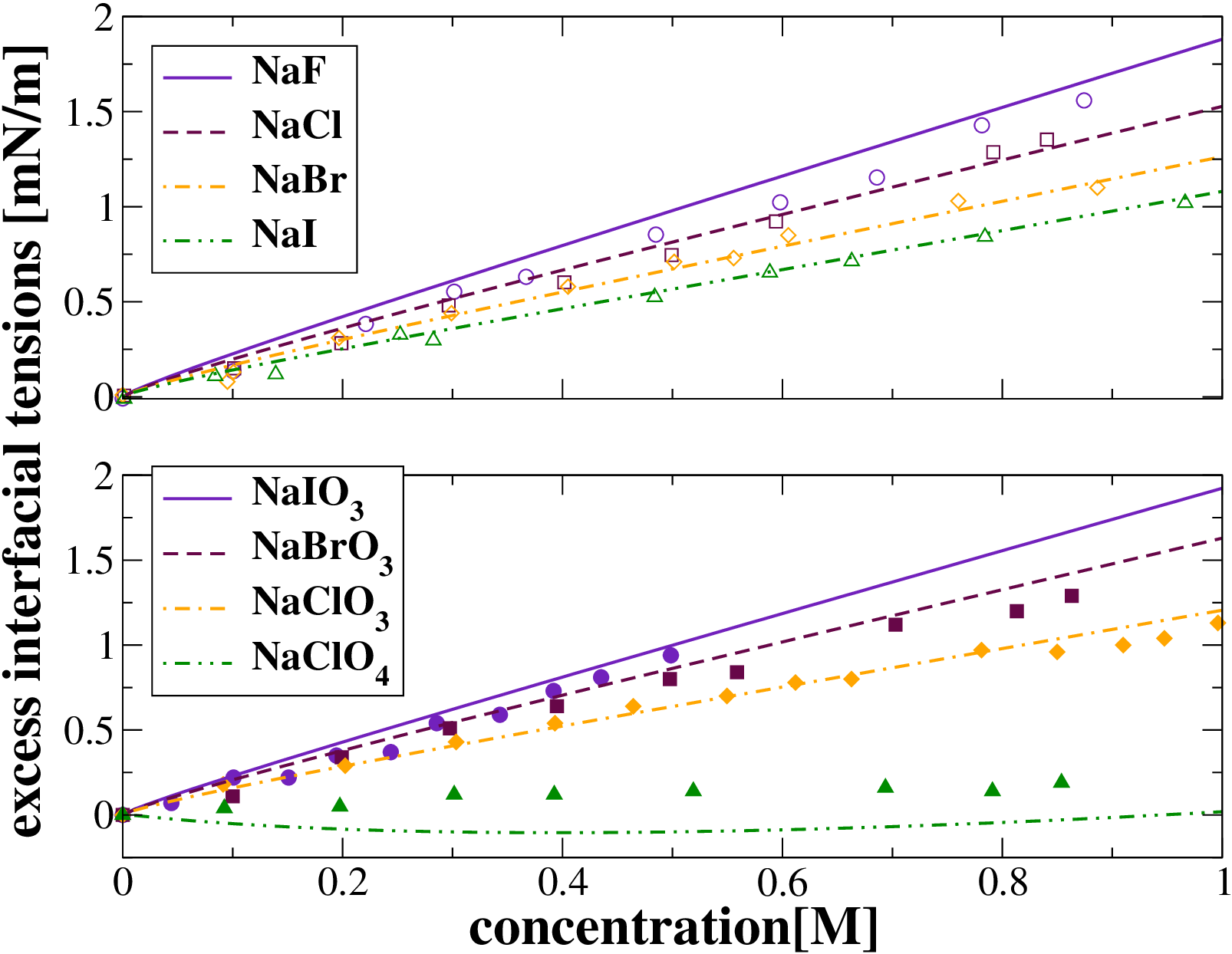}
  \caption{Excess interfacial tensions for various salts at the electrolyte-air
interface. The open circles, squares, diamonds and triangles represent
experimental data\cite{MaTs01,MaMa99,Ma} for \ce{NaF}, \ce{NaCl}, \ce{NaBr} and
\ce{NaI}, respectively. The full circles, squares, diamonds and triangles
represent experimental data\cite{Ma} for \ce{NaIO3}, \ce{NaBrO3}, \ce{NaClO3}
and \ce{NaClO4}, respectively. The lines represent the present theory.}
  \label{fig1}
\end{figure}

The partially hydrated  radius of the sodium cation
\ce{Na+} is the only free adjustable parameter of the theory.
It is obtained by fitting the surface tension of the \ce{NaI}
solution.  The calculation is performed by  first numerically solving the 
PB equation to obtaining the bulk
concentration of electrolyte $\rho_{\pm}(0)$ at the center of the 
drop. Then the Gibbs adsorption
isotherm 
eqn~(\ref{gaie}) is integrated numerically to calculate the 
excess surface tension of the electrolyte solution.  We find that  
$a=2.5$~\AA\ for \ce{Na+} gives an excellent fit to the experimental data~\cite{LeDo09},
Fig.~\ref{fig1}.  The same radius of \ce{Na+} is then used to
calculate the excess surface
tensions of other electrolyte solutions.
Note that while \ce{Br-} is a chaotrope, both \ce{F-}
and \ce{Cl-} are kosmotropes.  Furthermore for \ce{F-} the JD viscosity
B-coefficient is large and positive while for \ce{Cl-} it is almost zero.
Therefore near the interface,  \ce{F-} will remain fully hydrated with
the effective
radius~\cite{Ni59} $a=3.52$~\AA, while \ce{Cl-} is very weakly hydrated,
with the radius $a=2$~\AA\ close to its crystallographic size. 
The calculated excess surface
tension for all halide salts are in excellent agreement with the experimental
measurements, Fig.~\ref{fig1}.

At the moment there is no general theory for ionic hydration.  
For halide anions we saw that there was a very good correlation between the
size of the ion, its JD viscosity B-coefficient, and the hydration characteristics near the air-water interface.  One might hope that
such correlations will also persist for more complicated anions as well.  This, however,
is not the case. For example, iodate, \ce{IO3-}, is a very large anion, yet its
JD viscosity B-coefficient is similar to that of fluoride.  
Indeed, calculating the surface tension of \ce{NaIO3} solution, we 
find that \ce{IO3-}
must be treated as a strongly kosmotropic anion.  This is also
consistent with the recent \textit{ab initio}
simulations of Baer~et~al.~\cite{BaPh11}. Although the correlation between
the ionic size and ionic hydration is lost for more complicated oxy-anions, the
correlation between the B-coefficient and hydration  persists.  
This correlation can, therefore, be used to distinguish 
between the kosmotropes and the chaotropes
in the case of more complex anions~\cite{DoDi10}.

We now consider salts with oxy-anions.  
Although oxy-anions are not spherical, their effective radii are 
well described by an empirical formula based on experimentally
measured entropies of hydration~\cite{CoLa57}, $a = \frac{
n_{oxy}}{4} \left(d+1.4 \text{\ \AA}\right)$, where $n_{oxy}$ is the number of
oxygens in the anion and $d$ is the halogen-oxygen covalent bond length in the
corresponding salt crystal~\cite{Ni59}. We first consider \ce{NaIO3} solution. The
partially hydrated radius of sodium is the same as before, $a=2.5$~\AA. 
The JD viscosity B-coefficient~\cite{JeMa95} of \ce{IO3-}   is 
large and positive, $0.14$. This means that  
iodate will remain fully hydrated near the interface, keeping its bulk hydration 
radius~\cite{Ni59} $a=3.74$~\AA.  The ion-interface potential 
of \ce{IO3-} is then
$U_-(z)=U_i(z)$, with a hardcore repulsion at  $z=3.74$~\AA\ from
the GDS. Calculating the excess surface tension for \ce{NaIO3}, we
find a good agreement with experiment, 
Fig.~\ref{fig1}.   The ion \ce{BrO3-} has the B-coefficient~\cite{JeMa95} near zero, $0.009$, 
similar to \ce{Cl-} so, once again, we will treat these ion as a kosmotrope with
a partially hydrated radius equal to its bare size $a=2.41$~\AA. 
The result of this calculation is shown in 
Fig.~\ref{fig1}. The ions \ce{ClO3-} and \ce{ClO4-} have negative B
coefficients~\cite{JeMa95}, $-0.022$ and $-0.058$, respectively, and are chaotropes. 
The total ion-interface interaction potential for these ions is
$U_-(z)=U_i(z)+U_p(z)+U_c(z)$, with bare ionic radii~\cite{CoLa57}, $a=2.16$ and
$a=2.83$~\AA; and relative polarizabilities~\cite{PyPi92,DoDi10},
$\alpha=0.52$ and $\alpha=0.24$, respectively. The calculated excess
interfacial tension for \ce{NaClO3} agrees very well with the
experimental data, Fig.~\ref{fig1}. The agreement is not very good for the sodium
perchlorate, Fig.~\ref{fig1}. This ion is very big, so that a small error in its
effective radius calculated using the empirical formula presented above 
leads to a large error in its cavitational free energy  --- cavitational energy
scales with the cube of the radius --- resulting in an incorrect estimate of adsorption.
The theory also allows us to estimate the electrostatic potential difference across
the interface~\cite{DoDi10}, $\phi(R)-\phi(0)$, which is reported in Table~\ref{tab1}
for 
various salts at $1$M concentration.  The results are in
qualitative agreement with the experimental measurements of Frumkin~\cite{Fr24} and 
Jarvis~and~Scheiman~\cite{JaSc68}. Furthermore, if the ions are arranged in the order of increasing surface 
potential, one finds precisely the celebrated Hofmeister series~\cite{DoDi10}.
%%%%%%%%%%%%%%%%% table %%%%%%%%%%%%%%%%%%%%
\begin{table}[h]
\small
\centering
\caption{Surface potentials difference at $1$ M for various salts}
\label{tab1}
\begin{tabular}{l c c c} %table columns can also be centre- or right-aligned using 'c' or 'r' instead of 'l' here
      \hline
       Salts       &    Calculated (mV)   &   Frumkin \cite{Fr24,Ra63} (mV)  &  
Jarvis et al. \cite{JaSc68} (mV)  \\
      \hline
      \ce{NaF}     &      4.7             &        --       &         --        
  \\
      \ce{NaCl}    &     -2.1             &        -1       &     $\approx$ -1  
  \\
      \ce{NaBr}    &     -9.4             &        --       &     $\approx$ -5  
  \\
      \ce{NaI}     &     -14.3            &       -39       &     $\approx$ -21 
  \\
      \ce{NaIO3}   &      5               &       --        &       --          
  \\
      \ce{NaBrO3}  &     -0.12            &       --        &       --          
  \\
      \ce{NaNO3}   &      -8.27           &       -17       &     $\approx$  -8 
  \\
      \ce{NaClO3}  &     -11.02           &       -41       &       --          
  \\
      \ce{NaClO4}  &     -31.1            &       -57       &       --          
  \\
      \ce{Na2CO3}  &      10.54           &         3       &     $\approx$   6 
  \\
      \ce{Na2SO4}  &      10.17           &         3       &     $\approx$   35
  \\
      \hline
   \end{tabular}
\end{table}
%%%%%%%%%%%%%% end of table %%%%%%%%%%%%%%%%

\section{Acids solutions.}

\begin{figure}
\centering
  \includegraphics[height=8cm]{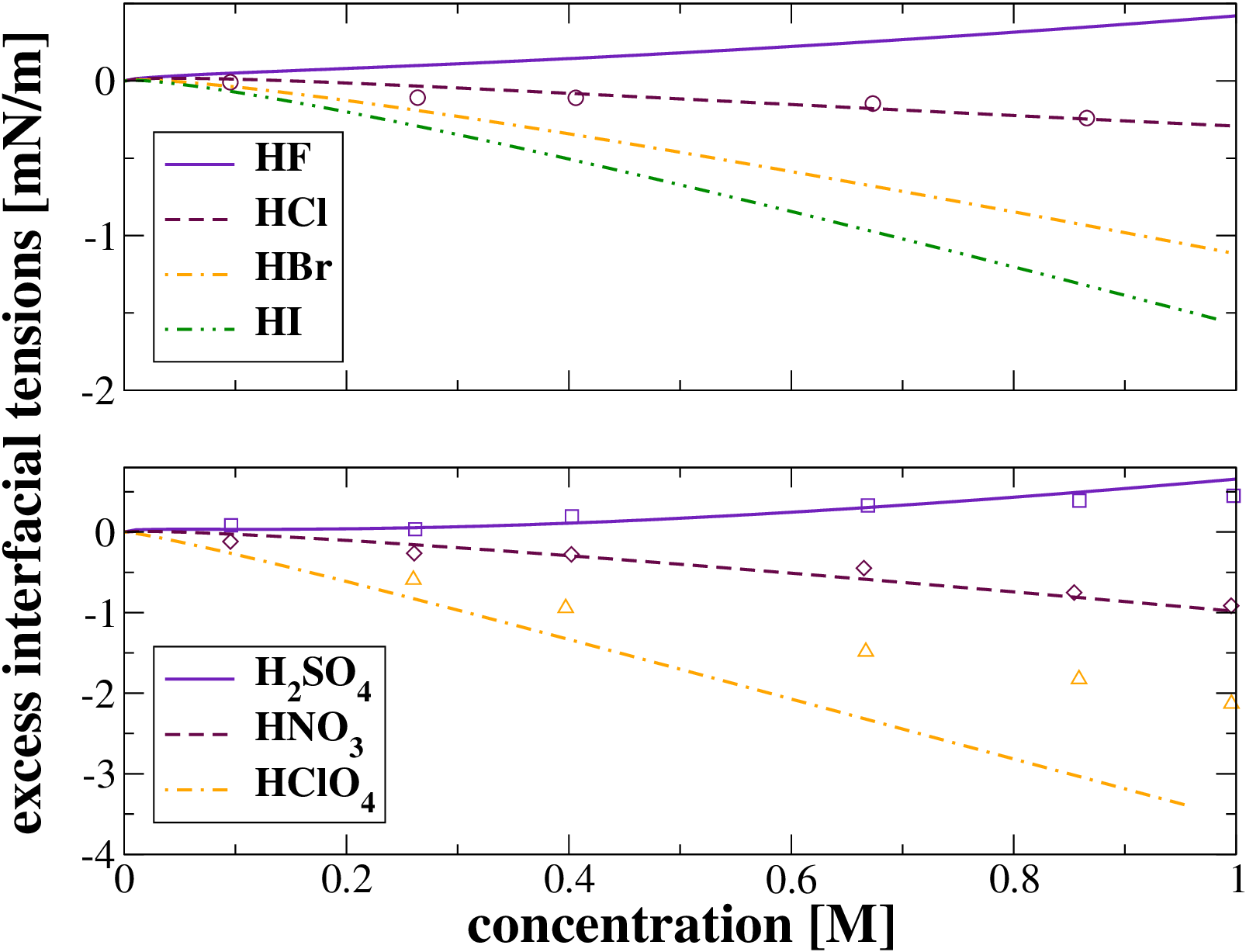}
  \caption{Excess interfacial tensions for various acids at the electrolyte-air
interface. The open circles, squares, diamonds and triangles represent
experimental data~\cite{WePu96} for \ce{HCl}, \ce{H2SO4}, \ce{HNO3} and
\ce{HClO4}, respectively. The lines represent the present theory.}
  \label{fig2}
\end{figure}

While most salts tend to increase the surface tension of the air-water interface
most acids do precisely the opposite. It is well known that proton \ce{H+} 
interacts strongly with the water molecules~\cite{Ei64,Zu00,MaTu99}, forming
complexes such as \ce{H3O+} and \ce{H2O5+}. The trigonal pyramidal
structure of 
hydronium~\cite{PeIy04,MuFr05} favors strong adsorption 
at the water-air interface, with  
the oxygen pointing towards the air~\cite{JuTo06,LeSi07,MuFr05,PeIy04,IyDa05}. 
For many acids the protonation of the interface is so strong as to result in a negative
excess surface tension.  To take this into account~\cite{DoLe10}
we add an additional adsorption potential for \ce{H+},
%%%%%%%%%%%%%%%%%%%%%%%%%%%%%%%%%%%%%%%%%%%
\begin{eqnarray}\label{uhtotal}
\beta U_h(z)=\left\{
\begin{array}{l}
0 \, \text{ for } z  \ge  1.97\ \text{\AA} \ , \\
-3.05 \, \text{ for } z < 1.97\ \text{\AA} \ .
\end{array}
\right.
\end{eqnarray}                                              
%%%%%%%%%%%%%%%%%%%%%%%%%%%%%%%%%%%%%%%%%%%%
where the value $-3.05$ was adjusted in order to obtain the correct excess
interfacial tension for the hydrochloric acid, Fig.~\ref{fig2}. The range of this
potential is taken to be $1.97$~\AA, the  length of the hydrogen bond. The anions
are treated as before --- classified as kosmotropes or
chaotropes~\cite{DoLe10} --- while the proton 
interacts with the interface through the potential $U_+(z)=U_i(z)+U_h(z)$. In the
image part of these potential the radius of proton is set to zero.

In Fig.~\ref{fig2} the excess interfacial tensions for
various acids are plotted.  The agreement with the experimental data is very
good for \ce{H2SO4} and \ce{HNO3}. As for sodium perchlorate,  \ce{HClO4} also 
shows a significant deviation from the experimental data, indicating again that
our estimate of the effective radius of \ce{ClO4-} is too large. 

To calculate the electrostatic potential difference across the interface
we integrate the PB equation.  This, however, will not account for the
reorientation of the 
interfacial water molecules when hydronium complex is formed, resulting in a formation
of dipole layer with a corresponding potential drop. As was discussed
earlier \ce{H3O+} prefers to orient itself with the hydrogens 
pointing towards the bulk water. The number
of hydroniums formed can be estimated from the 
proton adsorption, $N_+=\left[N - V
\rho_+(0) \right]$, where $\rho_+(0)$ is the 
bulk concentration, given by eqn~(\ref{pb}).  The internal electric field
inside the dipole layer can be estimated to be $E=4 \pi p N_+/\epsilon_o S d$, where
$p=0.38541 \ q$\AA\ is the water dipole moment, $d$ is the dipole length, and $S$ is
the interfacial area.  Integrating this field across the interface, we find
the contribution of the surface hydroniums to the overall potential difference
across the interface to be $\Delta \phi_w=-69604.5 \ \Gamma_+$, in mV.  Summing
this with the contribution arising from the PB equation, we obtain the overall potential
drop across the air-water interface, reported in  Table~\ref{tab2}.   The theoretical
results are also compared with the data of Frumkin~\cite{Fr24}.  
In view of the roughness of the estimates presented
above the qualitative agreement between the theory and experiment is quite reasonable.
%%%%%%%%%%%%%%%%% table %%%%%%%%%%%%%%%%%%%%
\begin{table}[b]
\small
\centering
\caption{Surface potential differences for various acids at the water-air interface, contributions from
electrolyte and aligned water dipoles}
\label{tab2}
\begin{tabular}{l c c } %table columns can also be center- or right-aligned using 'c' or 'r' instead of 'l' here
      \hline
      Acids                 &  calculated [mV]   & Frumkin~\cite{Fr24} [mV] \\
      \hline
      \ce{HF}            &        $85.5$      &         $-71$             \\
      \ce{HCl}            &       $1.24$      &         $-23$             \\
      \ce{HNO3}       &        $-84.4$     &         $-48$             \\
      \ce{HBr}          &        $-95$       &         $-34$             \\
      \ce{HI}            &         $-144.8$    &         $-61$             \\
      \ce{HClO4}      &        $-412$      &         $-82$             \\
      \hline
   \end{tabular}
\end{table}
%%%%%%%%%%%%%% end of table %%%%%%%%%%%%%%%%
The value $-71$mV for \ce{HF} reported in Ref.~\cite{Fr24} most like has as a 
wrong sign, since it falls completely outside the general trend. 

\section{Electrolyte-oil interface.}

The good agreement between the theory and experiments found above suggests
that the physical picture behind the mechanism of the ion-interface interaction
is more-or-less correct.  In particular we see that the ions near the air-water
interface must be divided into two classes: kosmotropes and chaotropes~\cite{DoDi10}.
While the kosmotropes remain hydrated near the interface the chaotropes
lose their hydration shell and, as the result of the hydrophobic cavitational
forces and high polarizability, become partially adsorbed at the interface.  
In the recent \textit{ab initio} simulations, Baer~and~Mundy~\cite{BaMu11} 
have calculated
the potential of mean force for 
iodide near the air-water interface, finding an almost perfect agreement
with the theory presented above~\cite{Le09}.  This suggests that 
the dispersion (van der Waals) interactions do not play a significant role at 
the air-water interface. To see why this might be the case 
let us first consider a
kosmotropic ion. Near the interface such ions remain hydrated,
interacting with almost the same number of water molecules as in the bulk, so
that dispersion contribution to their total free energy of solvation is
not affected by the presence of the interface.

For chaotropic ions absence of dispersion interactions is not so easily understood. 
It is possible, however, to make the following argument: most of the ionic 
charge of a chaotropic anion as it crosses the GDS concentrates in 
water, resulting in a large electric field~\cite{Le09}. This strong 
field attracts water molecules so that it is possible for a chaotropic ion
to interact dispersively with the same number of water molecules as it did in the bulk.  
To see if this argument is consistent, we will now study the effect of electrolyte
on the interfacial tension
of the oil-water interface.

Similar to what happens at 
the air-water interface, the kosmotropic ions near the oil-water 
interface will feel the ion-image interaction and the
hardcore repulsion from the GDS.  Oil, like air, has low dielectric 
constant, $\epsilon_o \approx 2$,
so that the ion-image and the polarization potentials, eqns~(\ref{Uim}) and (\ref{Upol}), will remain the same 
as at the air-water interface.
The chaotropic ions are driven towards the interface by their cavitational potential.
When part of the ion penetrates into oil, there is also a cavitation energy penalty from
the oil side.  The cavitational energy, is mostly entropic --- related to the number of water/oil molecules excluded from the cavity produced by the ion. 
Molecular weight of oil 
(dodecane used in the experiments) is
$10$ times higher while its mass density is the same as that of water. This
means that the number of exclude molecules in the ion cavity, and consequently
the cost of cavitational energy, in oil will be about $10$ times smaller 
than in water,
and can be safely neglected. Therefore, the cavitational potential
of a chaotrope at the water-oil
interface will remain the same as the air-water interface, eqn~(\ref{cavpot}). 

The dispersion potential should be proportional to the ionic
polarizability and the ionic volume exposed to oil.  We suggest the 
following simple phenomenological expression~\cite{DoLe12}
%%%%%%%%%%%%%%%%%%%%%%%%%%%%%%%%%%%%%%%%%%%
\begin{eqnarray}
\label{edis}
U_{d}(z)=\left\{
\begin{array}{l}
 0 \text{ for } z \ge  a  \ , \\
 A_{eff} \alpha \left[1 - \frac{(z/a + 1)^2(2 - z/a)}{4} \right] \text{ for }
-a<z<a \ ,
\end{array}
\right.
\end{eqnarray}
%%%%%%%%%%%%%%%%%%%%%%%%%%%%%%%%%%%%%%%%%%%%
where $A_{eff}$ is the effective Hamaker constant. Since at the oil-water
interface there is dispersive contribution to the adsorption potential,
the Hamaker constant for the oil-water interface should be
$A_{eff}\approx
A_{mw}^v - A_{mo}^v$, where $A_{mw}^v$ and  $A_{mo}^v$ are the metal-water 
and metal-oil~(dodecane) Hamaker constants in vacuum.
The metal constants are used since the ionic polarizability 
is already included in eqn~(\ref{edis}). Using the tabulated
values of the Hamaker constants, Ref.~\cite{HoWh80}, we obtain 
$A_{eff} \approx -4k_BT$.  Note that this is only a rough estimate of the strength
of the dispersion interaction.  In practice, we will adjust the value of $A_{eff}$
to obtain the measured interfacial tension of the \ce{KI} solution.

\begin{figure}[h]
\centering
  \includegraphics[height=8cm]{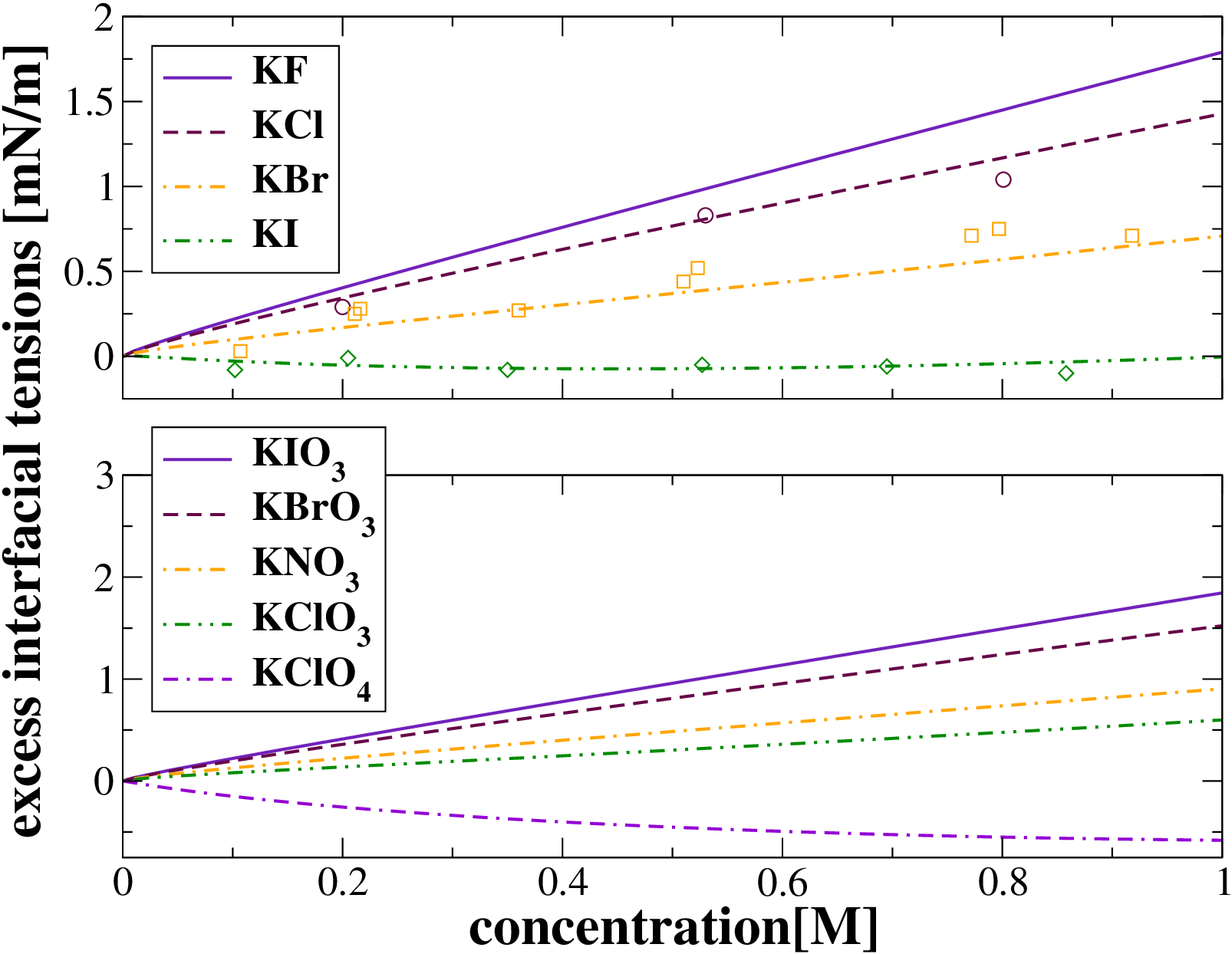}
  \caption{Excess interfacial tensions for various electrolyte solutions. The open circles, squares and diamonds represent experimental
data~\cite{AvSa76} for \ce{KCl}, \ce{KBr} and \ce{KI}, respectively. The lines
are calculated using the present theory.}
  \label{fig3}
\end{figure}

The interfacial tensions will be calculated as before.  We will solve the 
modified PB equation, eqn~(\ref{pb}), inside a drop, with the 
potentials $U_+(z)=U_i(z)$ for \ce{K+}, $U_-(z)=U_i(z)$ for
kosmotropes, and $U_-(z)=U_i(z)+U_p(z)+U_c(z)+U_d(z)$ for chaotropic anions.
From this solution we will calculate the ionic adsorption and,
integrating the Gibbs adsorption isotherm eqn~(\ref{gaie}), will obtain
the interfacial tensions. All the parameters used for kosmotropes and
chaotropes are the same as in the previous sections. 
The hydrated radius of the \ce{K+} is adjusted to obtain the experimentally
measured surface tension of  \ce{KCl} solution, Fig.~\ref{fig3}. 
We find that the potassium ion is partially hydrated with radius  of $a=2$~\AA. 
This radius will
be used for all the potassium salts. To obtain the effective Hamaker
constant for chaotropic ions, we  study the \ce{KI} solution, Fig.~\ref{fig3}. 
Fitting the experimental data we obtain 
$A_{eff}=-4.4\ k_B T$, which is in excellent agreement with our theoretical estimate,
suggesting that our physical picture about the role of dispersion interactions at
the air-water and oil-water interfaces is correct.
This Hamaker constant will be used for all the chaotropic anions. 
In Fig.~\ref{fig3}, we present the calculated 
interfacial tensions for various potassium salts. 
Unfortunately, the only additional experimental data available to us is for \ce{KBr},
which agrees very well with the predictions of the present theory.

\section{Acids-oil interface.}

\begin{figure}[h]
\centering
  \includegraphics[height=8cm]{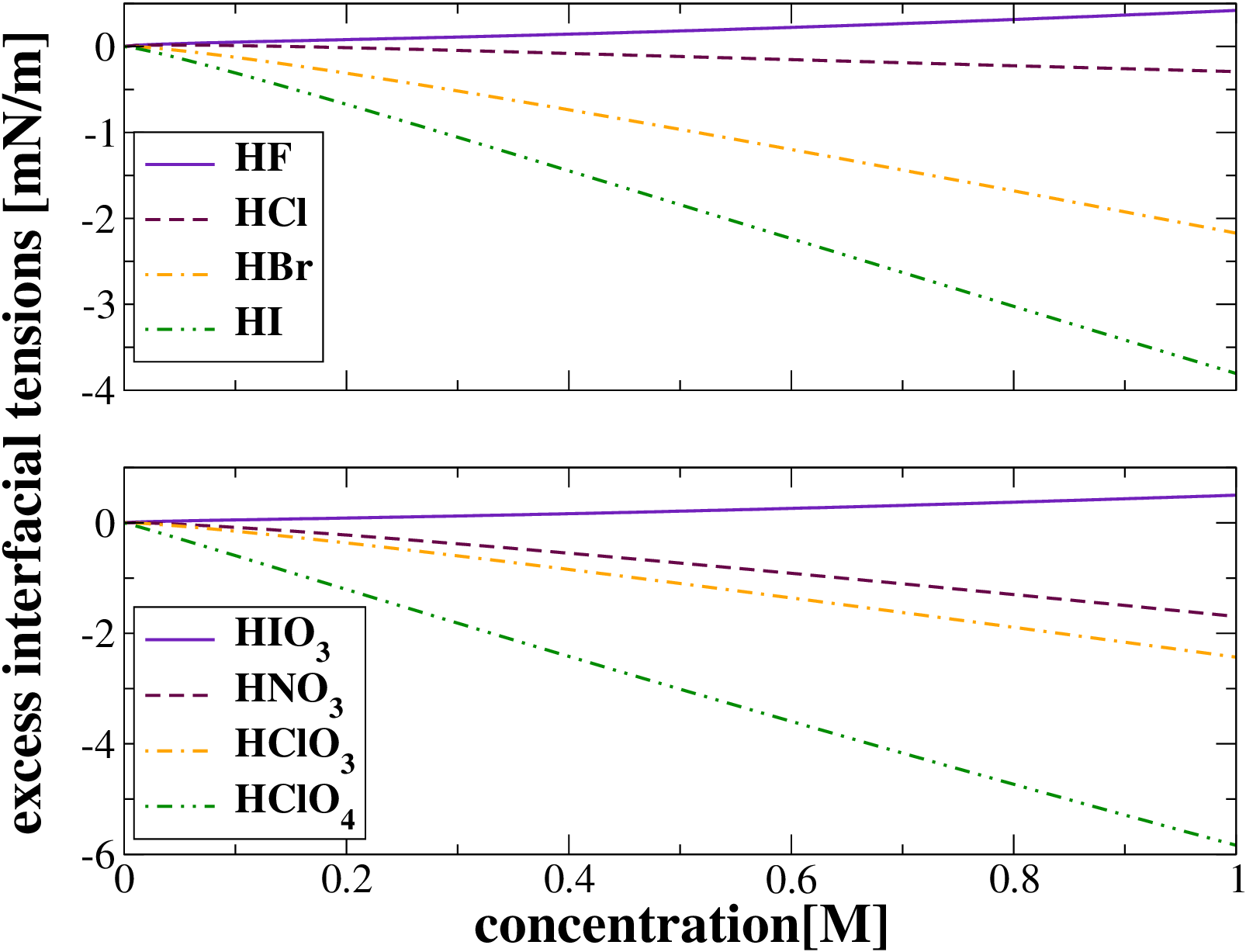}
  \caption{Excess interfacial tensions for different acids.  The lines are calculated using the present theory.}
  \label{fig4}
\end{figure}

We will now explore the effect of acids on the interfacial tension of
the water-oil interface.   It was shown previously that the 
hydronium ion \ce{H3O+} has a particular preference for the interfacial solvation.
This happens because the hydrogens of the hydronium ion are very good
hydrogen bond donors, while the oxygen is a bad receptor~\cite{PeIy04}.  This
leads to a 
preferential orientation of the hydronium ion at the water-air
interface, with the hydrogens pointing into the aqueous environment and the oxygen
sticking out. Calculations of solvation free energy confirm this interfacial 
behavior~\cite{Da03}. Here we will suppose that this basic picture 
persists for hydronium ion at the water-oil interface as well. Since at the moment
there is no experimental data on the interfacial tension of acid solutions
that can be used to re-parametrize our model, we will use the same
adsorption energy of proton as at the air-water interface~\cite{DoLe10},
-3.05~$k_BT$. The dispersion interaction and the cavitational potential
are also the same as used in the previous sections. Integrating the modified
PB equation and the
Gibbs adsorption isotherm, we obtain the ionic adsorptions and the excess interfacial
tensions of different acids.  In
Fig~\ref{fig4} we present our results. A significant decrease in the pure
water-oil interfacial tension is observed for acids containing 
chaotropic anions. Unfortunately at the moment there is
no experimental data available to test the predictions of the present theory.

The theory can also be used to estimate the electrostatic potential across the
acid-oil interface.  The calculation is analogous to the one performed
for the acid-air interface~\cite{DoLe10}. The results are presented in the 
Table~\ref{tab3}. 
It is very probable that the calculated potential differences are too large, since the theory
is not fully self-consistent.  Nevertheless the results provide us with a  
magnitude of the electrostatic potential difference that can be expected
across the water-oil interface for different acids.
%%%%%%%%%%%%%%%%% table %%%%%%%%%%%%%%%%%%%%
\begin{table}[h]
\small
\centering
\caption{Electrostatic potential differences for various acids}
\label{tab3}
\begin{tabular}{l c } %table columns can also be center- or right-aligned using 'c' or 'r' instead of 'l' here
      \hline
      Acids                 &  calculated [mV]    \\
      \hline
      \ce{HNO3}       &        $-154.53$         \\
      \ce{HBr}          &        $-196.67$         \\
      \ce{HClO3}      &        $-218.36$         \\
      \ce{HI}            &         $-320.65$         \\
      \ce{HClO4}      &        $-453.78$         \\
      \hline
   \end{tabular}
\end{table}
%%%%%%%%%%%%%% end of table %%%%%%%%%%%%%%%%

\section{Conclusions}

We have presented a general theory which allows us to calculate the surface
and the interfacial tensions of electrolyte solutions.  The theory
provides a very interesting picture of ionic specificity.  We find
that near air-water interface or a general hydrophobic surface 
ions can be divided into two classes, kosmotropes and chaotropes~\cite{DoLe11,DoDi10}.
Near the interface kosmotropes remain hydrated and are repelled from the
GDS.  On the other hand, chaotropes loose their hydration sheath and,
as a result of large polarizability, can become adsorbed to the hydrophobic 
interface. The theory also shows that the hydronium ion has
a strong preference for the interfacial solvation~\cite{DoLe10}. It is believed that the
surface water molecules are preferentially oriented with the hydrogens
sticking out towards the air.  To account for the measured surface potential
difference of acid solutions 
we find that the hydronium cation must orient itself opposite to the surface water, 
with its hydrogens pointing toward the bulk. 

Beyond a qualitative picture, the theory presented in this Faraday Discussion
paper  allows us to make
quantitative predictions about the surface tension and the electrostatic potential
of both electrolyte and acid solutions.  The theory can also be
extended to quantitatively calculate the critical coagulation concentrations
of hydrophobic colloidal suspensions, providing a new insight into 
the physical mechanisms responsible for the ionic specificity~\cite{DoLe11}.

Although the theory helps us to understand the physics behind the Hofmeister
series, there are still a number of issues that must be explored.  One of them
is the role of the surface potential of water.  The dielectric continuum theory
presented in this paper completely ignores the surface potential of water.  This is justified
{\it a posteriori} by the good agreement between the theory and the experimental measurements of surface and interfacial tensions 
of different acids and electrolytes.  Nevertheless, classical
point charge models~\cite{JoCh83,DaCh97,KaKu11} predict a surface potential of approximately $-600$mV.  If such
surface potential really exists, it should completely change the
electrostatics of ionic solvation, favoring a much stronger adsorption of the 
chaotropic anions than was found in the present theory.  
This, however, can not be true, since  this would
result in erroneous excess surface tensions of electrolyte 
solutions.  In fact, it has now been realized that
the polarizable force fields simulations, which have stimulated the development of the
present theory, predict too much adsorption of the chaotropic anions.  We speculate
that the reason for this excess of adsorption is precisely the artificial surface
potential of point charge water models.  The natural question to ask then is the following: if the existing 
classical models can not account for the surface properties of water, 
what about the full quantum mechanical calculations of the air-water interface?
In fact, the recent {\it ab initio} simulations~\cite{KaKu08,KaKu09,Le10} show that the surface potential of water
is not $-600$mV  but is $+3000$mV.  Note the difference in sign and the magnitude
of this potential!  This electrostatic potential difference across the  
air-water has been measured by high energy electron holography~\cite{KaKu11}.  
Nevertheless, the authors of Ref.~\cite{KaKu11},  argue that this huge potential  
is irrelevant for the electro-chemistry, in which case they suggest
the potential must be coarse grained on the scale of an ion.  If this is done properly,
they argue,
the surface potential of water felt by an ion such as \ce{I-} will drop to a few mV
and can be safely ignored.  
This is possibly the reason why the {\it ab initio} potential of mean force for \ce{I-} agrees 
so well~\cite{BaMu11} with the present dielectric continuum theory which completely neglects
the electrostatic surface potential of water.  More work is necessary to fully
elucidate these issues.

%\footnote[4]{Footnotes should appear here. These might include comments
%relevant to but not central to the matter under discussion, limited experimental
%and spectral data, and crystallographic data.} 

\section{Acknowledgements}

This work was partially supported by the CNPq, FAPERGS, INCT-FCx, and by the 
US-AFOSR under the grant FA9550-09-1-0283.

%If notes are included in your references you can change the title using the
%following command.
%\renewcommand\refname{Notes and references}
\bibliography{ref.bib} %your .bib file

\end{document}